\documentclass[prb,twocolumn,showpacs,superscriptaddress]{revtex4}%

\usepackage{mathrsfs}
\usepackage{amsfonts}
\usepackage{amsmath}
\usepackage{amssymb}
\usepackage{graphicx}%
\begin{document}
\title{Quantum Corrals and Quantum Mirages on the Surface of a Topological Insulator}
\author{Zhen-Guo Fu}
\affiliation{SKLSM, Institute of Semiconductors, CAS, P. O. Box 912, Beijing 100083, China}
\affiliation{LCP, Institute of Applied Physics and Computational Mathematics, P. O. Box
8009, Beijing 100088, China}
\author{Ping Zhang}
\thanks{Corresponding author: zhang\_ping@iapcm.ac.cn}
\affiliation{LCP, Institute of Applied Physics and Computational Mathematics, P. O. Box
8009, Beijing 100088, China}
\author{Zhigang Wang}
\affiliation{LCP, Institute of Applied Physics and Computational Mathematics, P. O. Box
8009, Beijing 100088, China}
\author{Shu-Shen Li}
\affiliation{SKLSM, Institute of Semiconductors, CAS, P. O. Box 912, Beijing 100083, China}

\begin{abstract}
We study quantum corrals on the surface of a topological insulator (TI).
Different resonance states induced by nonmagnetic (NM), antiferromagnetic
(AFM), and ferromagnetic (FM) corrals are found. Intriguingly, the spin is
clearly energy-resolved in a FM corral, which can be effectively used to
operate surface carrier spins of TI. We also show that an observable quantum
mirage of a magnetic impurity can be projected from the occupied into the
empty focus of a FM elliptic corral, while in NM and AFM corrals the mirage
signal becomes negligibly weak. In addition, the modulation of the interaction
between two magnetic impurities in the quantum corrals is demonstrated. These
prominent effects may be measured by spin-polarized STM experiments.

\end{abstract}

\maketitle

\section{ Introduction}

Spintronics and quantum information are active areas in the modern condensed
matter physics. Spin-orbit interaction (SOI) has opened a door for controlling
electron spins in materials, and some semiconducting spintronics devices based
on the SOI have been proposed in last decade \cite{Baibich}. Very recently, a
new quantum matter phase, named as topological insulator (TI) suggested by a
series of theoretical \cite{Kane, Bernevig,Fu,Moore, Qi,Zhanghj} reports, has
been detected experimentally in two-dimensional HgTe quantum well \cite{Konig}
and three-dimensional Bi$_{2}$Se$_{3}$-family materials \cite{Hsieh1,Chen,
Xia} characterized by the inherent strong SOI. One significant issue in such
systems is that the strong SOI-induced Berry phase (chiral spin texture)
\cite{Hsieh2,Roushan} suppresses backscattering, such that the topological
surface states exhibit quantum-entanglement effects \cite{Fu, Konig, Hsieh1}
and could be useful for performing fault-tolerant quantum computation
\cite{Zhang,Fu1,Moore1}.

In addition, some nanoscale atomic structures, in which the electrons are
confined, have potential applications in spintronics and quantum computations.
For example, quantum corrals, which is a closed geometrical array constructed
by impurities on tow-dimensional (2D) surface of materials, is a candidate for
applications in quantum information \cite{Kane1,Morr}. Especially, when a
manipulated magnetic atom is located at one focus of the elliptic corral, one
can detect the signature not only at this focus but also at the empty one,
which is named as quantum mirage effect. Therefore, the quasiparticle
interference state around impurities is an important candidate for these
potential applications of TI materials. The advanced progress of scanning
tunneling microscopy (STM) has made it possible to probe impurity sates for
various materials, and many efforts have been devoted to this active field.
Over the past few years, quantum corrals on surfaces of noble metals
\cite{Heller,Manoharan, Aligia, Hallberg,Moon,Rossi,Fransson,Barr,Walls} and
superconductors \cite{Morr} have been extensively studied for exploring the
surface electron interference effects. Lots of intriguing quantum phenomena
have been found, including quantum mirages \cite{Morr,Manoharan,Hallberg,
Aligia,Moon}, Kondo effect \cite{Rossi}, quantum invisibility \cite{Fransson},
electron lifetime and resonance widths \cite{Barr}, SOI effect in quantum
corrals \cite{Walls} and so on. Inspired by these interesting observations in
corrals and the potential applications to quantum computations of TI, in this
paper we study the quantum corrals, quantum mirages, and interactions between
two magnetic impurities located in corrals on the TI surface.\textit{
}Intriguingly, some unique phenomena are observed, which dramatically
distinguish the present corral-TI system from previously reported ones
harnessed on the ordinary metals or semiconductor heterostructures. We show
that an energy gap is opened in the ferromagnetic (FM) corral, while it is not
found in antiferromagnetic (AFM) and nonmagnetic (NM) cases. Remarkably, the
spin can be energy-resolvable (-unresolvable), and the quantum mirage is clear
(unclear) in FM (AFM) corrals. In addition, the role of NM and FM corrals in
modulating the interaction between two magnetic impurities is more prominent
than AFM ones. Our predictions, which can be measured using spin-polarized STM
\cite{Meier}, provide to reveal peculiar topological quantum interference
properties of TI surface states, based on which new applications for
spintronics and quantum computations with TI systems may be proposed.

\section{Model and method}

We model the TI surface, on which adatoms are adsorbed to construct a quantum
corral, by a low-energy effective massless Dirac Hamiltonian written as
\begin{equation}
H\mathtt{=}v_{F}\left(  \boldsymbol{k}\mathtt{\times}\boldsymbol{\sigma
}\right)  \mathtt{\cdot}\hat{z}\mathtt{+}\sum_{i=1}^{N}V_{i}\left(
\boldsymbol{r}\right)  ,
\end{equation}
where $v_{F}$ is the Fermi velocity, $\boldsymbol{k}\mathtt{=}\left(
k_{x},k_{y}\right)  $ is the planar momentum, $\boldsymbol{\sigma}%
\mathtt{=}\left(  \sigma_{x},\sigma_{y},\sigma_{z}\right)  $ are the Pauli
spin matrices, and $N$ is the total number of impurities. We take Fermi
velocity $v_{F}\mathtt{=}1$, and use lattice constant $a_{0}\mathtt{=}1$ as
the distance unity and $E_{0}\mathtt{=}v_{F}/a_{0}$ as the energy unity.
\begin{equation}
V_{i}\left(  \boldsymbol{r}\right)  \mathtt{=}(U_{i}\sigma_{0}\mathtt{+}%
\frac{1}{2}J_{i}\boldsymbol{S}\mathtt{\cdot}\boldsymbol{\sigma})\delta\left(
\boldsymbol{r}\mathtt{-}\boldsymbol{r}_{i}\right)
\end{equation}
is the potential of the $i^{\text{th}}$ impurity located at site
$\boldsymbol{r}_{i}$, with $U_{i}$ ($J_{i}$) the potential (magnetic)
scattering strength, and $\boldsymbol{S}$ the spin of a magnetic impurity. We
define $U_{i}$ ($J_{i}\mathtt{=}0$) for scalar impurities, while $J_{i}$
($U_{i}\mathtt{=}0$) for magnetic ones in all calculations. If the
measurements are performed at a temperature higher than the Kondo temperature,
which is $\mathtt{\sim}30$ K in the magnetic doped Bi$_{2}$Se$_{3}$
nanoribbons \cite{Cui}, the coupling $J_{i}$ will not exceed the critical
value $J_{cr}$ before a Kondo effect occurs. For magnetic corrals, in this
paper we assume that the distance between impurities is large enough and the
exchange coupling $J_{i}\mathtt{\ll}J_{cr}$, so that the
Ruderman-Kittel-Kasuya-Yosida interactions between impurity spins and Kondo
screening of the spin $\boldsymbol{S}$ by the band electrons are neglected,
and the impurity spin acts as a classical local magnetic moment $M_{0}%
\mathtt{=}J_{i}S/2$ under mean-field approximation
\cite{Morr,Liu,Biswas,Balatsky}.

The quantum interference effects that we discuss can be observed in
measurements of the TI surface electron local density of states (LDOS)
$\rho\left(  \boldsymbol{r},E\right)  $, which can be calculated by
\begin{equation}
\rho\left(  \boldsymbol{r},E\right)  \mathtt{=}\mathtt{-}\frac{1}{\pi
}\operatorname{Im}\left\{  \mathtt{Tr}\left[  G\left(  \boldsymbol{r}%
,\boldsymbol{r},E\right)  \right]  \right\}  .
\end{equation}
The electron Green's function is expressed as
\begin{align}
G\left(  \boldsymbol{r},\boldsymbol{r}^{\prime},E\right)   &  =G_{0}\left(
\boldsymbol{r},\boldsymbol{r}^{\prime},E\right)  +\sum\nolimits_{i,j=1}%
^{N}G_{0}\left(  \boldsymbol{r},\boldsymbol{r}_{i},E\right) \nonumber\\
&  \times T\left(  \boldsymbol{r}_{i},\boldsymbol{r}_{j},E\right)
G_{0}\left(  \boldsymbol{r}_{j},\boldsymbol{r}^{\prime},E\right)  ,
\end{align}
where the $T\mathtt{-}$matrix is%
\begin{equation}
T\left(  \boldsymbol{r}_{i},\boldsymbol{r}_{j},E\right)  =V_{i}\delta
_{i,j}+V_{i}\sum\nolimits_{l=1}^{N}G_{0}\left(  \boldsymbol{r}_{i}%
,\boldsymbol{r}_{l},E\right)  T\left(  \boldsymbol{r}_{l},\boldsymbol{r}%
_{j},E\right)  ,
\end{equation}
where the unperturbed Green's function $G_{0}\left(  \boldsymbol{r}%
,\boldsymbol{r}^{\prime},E\right)  $ can be obtained by the Fourier
transformation of the Green's function
\begin{equation}
G_{0}\left(  \boldsymbol{k},iE\right)  \mathtt{=}\left[  iE\mathtt{-}%
v_{F}\left(  \boldsymbol{k}\mathtt{\times}\boldsymbol{\sigma}\right)
\mathtt{\cdot}\hat{z}\right]  ^{-1}.
\end{equation}

For numerical calculations, the total Green's function $G\left(
\boldsymbol{r},\boldsymbol{r}^{\prime},E\right)  $ in Eq. (4) can be rewritten
as
\begin{equation}
G\left(  \boldsymbol{r},\boldsymbol{r}^{\prime},E\right)  =G_{0}\left(
\boldsymbol{r},\boldsymbol{r}^{\prime},E\right)  +\mathcal{G}_{0}%
\mathcal{TG}_{0}^{\prime},
\end{equation}
where%
\begin{align}
\mathcal{G}_{0} &  =\left(
\begin{array}
[c]{cccc}%
G_{0}\left(  \boldsymbol{r},\boldsymbol{r}_{1},E\right)   & \cdots & \cdots &
G_{0}\left(  \boldsymbol{r},\boldsymbol{r}_{N},E\right)
\end{array}
\right)  _{2\times2N},\\
\mathcal{G}_{0}^{\prime} &  =\left(
\begin{array}
[c]{c}%
G_{0}\left(  \boldsymbol{r}_{1},\boldsymbol{r}^{\prime},E\right)  \\
G_{0}\left(  \boldsymbol{r}_{2},\boldsymbol{r}^{\prime},E\right)  \\
\vdots\\
G_{0}\left(  \boldsymbol{r}_{N},\boldsymbol{r}^{\prime},E\right)
\end{array}
\right)  _{2N\times2},
\end{align}
\begin{figure}[ptb]
\begin{center}
\includegraphics[width=1.\linewidth]{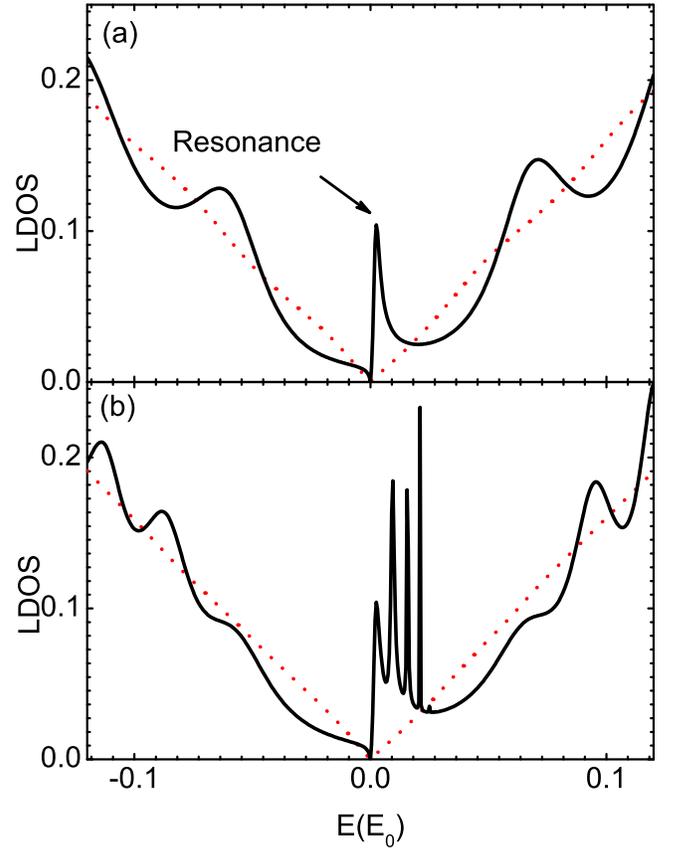}
\end{center}
\caption{ (Color online) $\rho\left(  \boldsymbol{r},E\right)  $ as a function
of $E$ at (a) $\boldsymbol{r}\mathtt{=}(0,0)$ and (b) $\boldsymbol{r}%
\mathtt{=}(25,0)$ in a NM circular corral with radius $R_{0}\mathtt{=}50$. The
red dashed and black curves are for $U\mathtt{=}0.36$ and $12.0$,
respectively.}%
\label{fig1}%
\end{figure}which denote the propagation of electrons from the STM tip to the
impurities as well as from the impurities to the STM tip. The
$2N\mathtt{\times}2N$ matrix $\mathcal{T}$ in Eq. (6) is a rewritten form of
the Eq. (5) for the $T\mathtt{-}$matrix, which is given by%
\begin{equation}
\mathcal{T}=\mathcal{H}_{imp}+\mathcal{H}_{imp}\mathbf{G}_{0}\mathcal{T}%
\end{equation}
with the impurity Hamiltonian
\begin{equation}
\mathcal{H}_{imp}=\left(
\begin{array}
[c]{cccc}%
V_{1} & \mathbf{0} & \cdots & \mathbf{0}\\
\mathbf{0} & V_{2} & \cdots & \mathbf{0}\\
\vdots & \vdots & \ddots & \vdots\\
\mathbf{0} & \mathbf{0} & \cdots & V_{N}%
\end{array}
\right)  ,
\end{equation}
containing all of the impurity scattering potentials $V_{i}$ ($i\mathtt{=}%
1,\cdots,N$). The information about the propagation between the impurities are
included in $\mathbf{G}_{0}$, which can be expressed as
\begin{equation}
\mathbf{G}_{0}=\left(
\begin{array}
[c]{cccc}%
G_{0}\left(  \boldsymbol{r}_{1},\boldsymbol{r}_{1}\right)   & G_{0}\left(
\boldsymbol{r}_{1},\boldsymbol{r}_{2}\right)   & \cdots & G_{0}\left(
\boldsymbol{r}_{1},\boldsymbol{r}_{N}\right)  \\
G_{0}\left(  \boldsymbol{r}_{2},\boldsymbol{r}_{1}\right)   & G_{0}\left(
\boldsymbol{r}_{2},\boldsymbol{r}_{2}\right)   & \cdots & G_{0}\left(
\boldsymbol{r}_{2},\boldsymbol{r}_{N}\right)  \\
\vdots & \vdots & \ddots & \vdots\\
G_{0}\left(  \boldsymbol{r}_{N},\boldsymbol{r}_{1}\right)   & G_{0}\left(
\boldsymbol{r}_{N},\boldsymbol{r}_{2}\right)   & \cdots & G_{0}\left(
\boldsymbol{r}_{N},\boldsymbol{r}_{N}\right)
\end{array}
\right)  .
\end{equation}
As a result, one can obtain
\begin{equation}
\mathcal{T}=\frac{\mathcal{H}_{imp}}{I-\mathcal{H}_{imp}\mathbf{G}_{0}},
\end{equation}
where $I$ is a $2N\mathtt{\times}2N$ unity matrix. Finally, by using the Eqs.
(6-12), one can easily finish the numerical calculations, and then get the LDOS.

\section{Results and discussions}

We first study a circular quantum corral of radius $50$ comprising $80$ NM
adatoms on TI surface. Comparing to the case of clean surface, we find that
the local correction of LDOS by the NM corral is weak when the potential is
not strong enough (such as $U\mathtt{=}0.36$), as indicated by the red dashed
curves in Fig. \ref{fig1}. With increasing the potential strength, however,
there turns to occur an impurity resonance peak projected at the Dirac point,
as shown by the black curves in Fig. \ref{fig1}, corresponding to a
sufficiently strong potential ($U\mathtt{=}12.0$). Furthermore, due to the
quantum confinement and interference induced by multiple scattering,
additional peaks at other band regions are found (see the black curve in Fig.
\ref{fig1}(b) for site $\boldsymbol{r}\mathtt{=}\left(  25,0\right)  $). This
prominently differs from the case of a single NM impurity \cite{Biswas} on the
TI surface. The NM resonant state at the Dirac point is doubly degenerate
according to the Kramers' theorem \cite{Balatsky}, and these peaks will be
located at the negative side for a negative potential $U$. Clearly, although
the resonance states appear in low-energy regime, signatures of fundamental
destruction of the Dirac point due to the NM corral, such as an energy gap
opening, does not occur. \begin{figure}[ptb]
\begin{center}
\includegraphics[width=1.\linewidth]{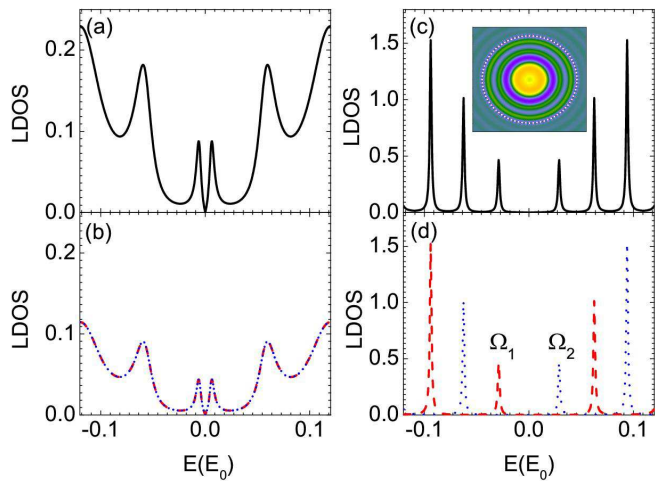}
\end{center}
\caption{(Color online) $\rho\left(  \boldsymbol{r}\mathtt{=}\mathbf{0}%
,E\right)  $ as a function of $E$ in an AFM (left panel) and a FM (right
panel) in circular corral with radius $R_{0}\mathtt{=}50$. In (a) and (c), the
black curves are for total LDOS. In (b) and (d), the blue dotted (red dashed)
curves correspond to the spin-up (spin-down) component of LDOS. Inset in (c):
spatial distribution of LDOS for a FM corral with $E\mathtt{=}0.32$. We chose
$M_{0}\mathtt{=}40.0$ ($M_{0}\mathtt{=}5.0$) for AFM (FM) corral in
calculations.}%
\label{fig2}%
\end{figure}

The time-reversal symmetry will be broken by magnetic impurities (we assume
the impurity spins to be polarized along the $\hat{z}$-axis), therefore, the
LDOS changes considerably when the quantum corrals are constructed by magnetic
impurities. For this straightforward reason, we analyze the following two
cases: (i) an AFM aligned corral with a zero total spin and (ii) a FM aligned
corral. Theoretical simulations of LDOS in the presence of an AFM circular
corral on TI surface are plotted in Fig. \ref{fig2}(a) and \ref{fig2}(b). It
is clear that due to the breaking of the time-reversal symmetry, differing
from the NM case, the resonance in the center of magnetic corrals splits into
spin-polarized peaks symmetrically located about both sides of the Dirac cone.
In fact, when the STM tip is moved from the center of corrals (i.e.,
$\boldsymbol{r}\mathtt{\neq}0$), these resonance states will split even more,
which is not shown here for briefness. It is found that by tuning the impurity
potential up to $M_{0}\mathtt{=}5.0$ the resonant states are not clear in the
AFM corral, therefore, $M_{0}\mathtt{=}40.0$ is representatively chosen in
Fig. \ref{fig2}(a) and \ref{fig2}(b). In the AFM case, at least two facts
should also be noticed. For one thing, no gap-opening at the Dirac point is
found because the total spin is zero in an AFM corral. For another, the
density of spin-up and spin-down states,
\begin{equation}
\rho_{\uparrow\left(  \downarrow\right)  }\left(  \boldsymbol{r},E\right)
\mathtt{=}\mathtt{-}\operatorname{Im}G_{11\left(  22\right)  }\left(
\boldsymbol{r},E\right)  /\pi,
\end{equation}
are energy-unresolvable in AFM circular corral, as shown in Fig. \ref{fig2}(b).

Through comparison of the numerical results between AFM and NM corrals, we can
conclude that the role of magnetic impurities in reshaping low-energy LDOS
should be more subtle than NM scattering potentials on TI surfaces, which is
consistent with the results in previous single-impurity studies (e.g. Refs.
\cite{Liu,Biswas}). This argument is more remarkable in FM corrals, in which
the resonance states are easily obtained even with a weaker scattering
potential. This can be seen clearly in Fig. \ref{fig2}(c) and \ref{fig2}(d),
where the scattering strength is chosen as $M_{0}\mathtt{=}5.0$. The quantum
interference becomes so strong that not only the resonance peaks become
sharper, but also the LDOS is suppressed, and especially a gap is opened over
the Dirac point. This prominent phenomenon was not observed in both NM and AFM
corrals. Surprisingly, observing the Fig. \ref{fig2}(d) where the blue dotted
curve is for $\rho_{\uparrow}\left(  \boldsymbol{r}\mathtt{=}0,E\right)  $
while red dashed one represents $\rho_{\downarrow}\left(  \boldsymbol{r}%
\mathtt{=}0,E\right)  $, we can find that the spin is energy-resolved in FM
corrals. For example, consider the pair of resonance peaks $\Omega_{1\left(
2\right)  }$ marked in Fig. \ref{fig2}(d). In this case, $\Omega_{1}$
($\Omega_{2}$) mainly arises from impurity scattering with spin-down (-up)
electrons, which could be measured in the spin-polarized STM experiments.
These pure spin-resolved states allows us to operate the spin-up or -down
electrons selectively, thus this is an excellent candidate for applications in
spin switch and magnetic storage element in nanometer scale. A typical spatial
oscillation pattern of the LDOS for a FM corral with energy close to the
experimentally accessible Fermi energy is shown in the inset of Fig.
\ref{fig2}(c).

\begin{figure}[ptb]
\begin{center}
\includegraphics[width=1.\linewidth]{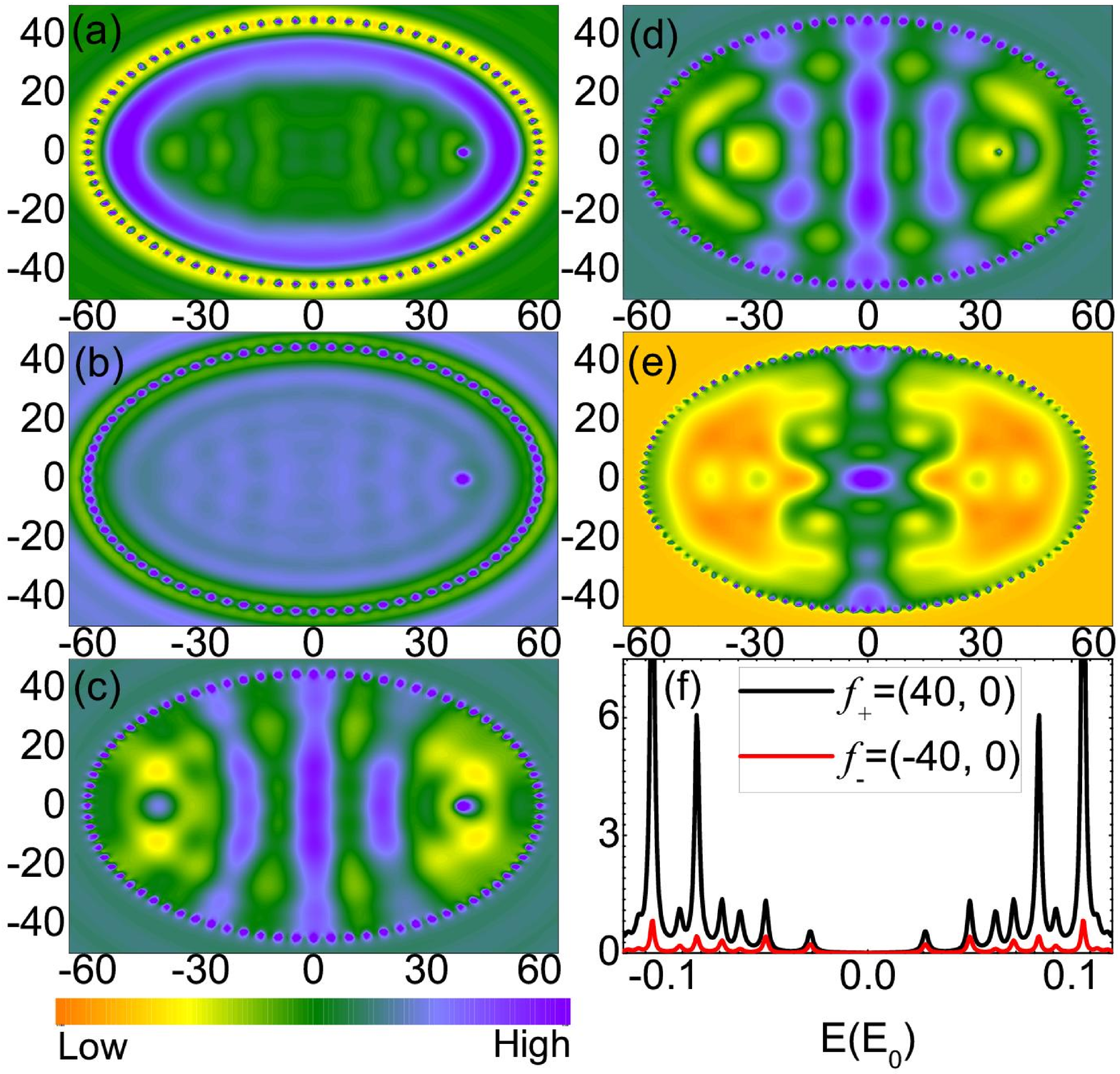}
\end{center}
\caption{ (Color online) $\rho\left(  \boldsymbol{r},E\right)  $ with
$E\mathtt{=}0.32$ for a magnetic impurity located at $f_{+}\mathtt{=}(40,0)$
in (a) NM, (b) AFM, and (c) FM corrals, respectively. (d), (e) LDOS
$\rho\left(  \boldsymbol{r},E\right)  $ for a magnetic impurity located at
$\boldsymbol{r}\mathtt{=}(35,0)$ and $\boldsymbol{r}\mathtt{=}(0,0)$ in FM
corrals, respectively. Also, (f) LDOS in the occupied (black curve) and empty
focus (red curve) for FM corrals with a magnetic impurity located at $f_{+}$
corresponding to (c). $U\mathtt{=}5.0$ and $M_{0}$=$5.0$ were used for
non-magnetic and magnetic corrals, respectively.}%
\label{fig3}%
\end{figure}

We turn now to focus on the quantum mirage effect of the elliptical corrals on
TI surface. In Fig. \ref{fig3} we present the quantum mirages in NM [Fig.
\ref{fig3}(a)], AFM [Fig. \ref{fig3}(b)], and FM [Fig. \ref{fig3}(c)-(f)]
corrals with $80$ impurities, $a\mathtt{=}60$, $b\mathtt{=}45$, eccentricity
$e\mathtt{=}\sqrt{7}/4$ and foci $f_{\pm}\mathtt{=}\left(  \mathtt{\pm
}40,0\right)  $. Usually, a good mirage effect is projected in certain
ellipses that have large peak amplitudes at the foci \cite{Fiete}. However, we
find that by tuning the impurity potential up to $U\mathtt{\ }$(or $M_{0}%
$)$\mathtt{=}5.0$ the resonant states are not clear in either NM or AFM
corrals, and thereby the quantum mirages in LDOS at empty focus $f_{-}$
projected by a magnetic impurity located at focus $f_{+}$ are too weak to be
observed as seen in Figs. \ref{fig3}(a) and \ref{fig3}(b). Things will be
absolutely different in a FM corral since the LDOS is suppressed and resonant
states could be easily obtained. As expected, a quantum mirage at the empty
focus, $f_{-}$ in a FM corral can be clearly observed [as seen in the spatial
pattern in the LDOS shown in Fig. \ref{fig3}(c)]. Corresponding to Fig.
\ref{fig3}(c), we show the spectral property of mirage versus the energy $E$
in Fig. \ref{fig3}(f), where the positions of resonance peaks of the mirage at
the empty focus $f_{-}$ (red curve) are in good agreement with those at the
occupied focus $f_{+}$ (black curve). Consequently, the FM quantum corrals
rather than NM and AFM ones can provide a good nanoscale facility to project
quantum mirages on the TI surface. The LDOS pattern will change when the
magnetic impurity is moved off the focus, similar to other systems
\cite{Morr}, since different locations of the impurity give rise to different
excited eigenmodes in corrals. For example, when the impurity is moved to
$\mathbf{r}\mathtt{=}\left(  35,0\right)  $, a much weaker quantum mirage is
projected at $f_{-}$, as illustrated in Fig. \ref{fig3}(d), and the spatial
oscillations in LDOS changes completely when the impurity is placed at the
center [$\mathbf{r}\mathtt{=}\left(  0,0\right)  $] as shown in Fig.
\ref{fig3}(e). Besides, the LDOS in the corrals with oscillating potentials
will exhibit some new features, and the quantum mirage will be destroyed (not
shown here). The quantum mirages in elliptical FM corral on TI surface can be
used to transmit information without wires, which offers a simple technique
for manipulating the coherent quantum-state in real-space \cite{Moon}, and may
open a new door for the technological realization of quantum computers of low
energy consumption by using TI materials.

\begin{figure}[ptb]
\begin{center}
\includegraphics[width=1.\linewidth]{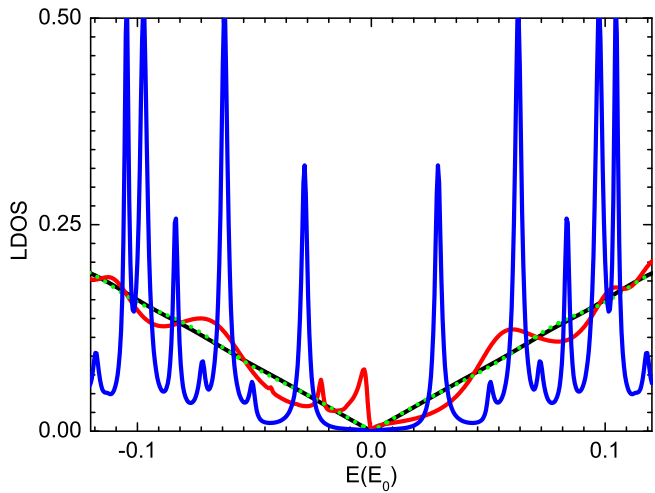}
\end{center}
\caption{ (Color online) Quantum interference in the LDOS $\rho\left(
\boldsymbol{r},E\right)  $ at $\boldsymbol{r}\mathtt{=}0$ generated from two
spin-parallel magnetic impurities located at $f_{\pm}\mathtt{=}\left(
\pm40,0\right)  $, where the black curve represents $\rho\left(
\boldsymbol{r},E\right)  $ in the absence of a corral. The red, green, and
blue curves denote the cases of NM, AFM, and FM corrals, respectively. We
chose $U\mathtt{=}M_{0}\mathtt{=}5.0$ in calculations.}%
\label{fig4}%
\end{figure}

The interference between two magnetic impurities is another interesting and
important issue, which is dependent on many physical factors including spatial
locations, impurity scattering strength, relative angle between two impurities
spins, and so on. We briefly discuss the influence of quantum corrals on the
quantum interference effects between two spin-parallel magnetic impurities
apart $2\left\vert f_{+}\right\vert $ on the TI surface. Taking into account
these two magnetic impurities, and using the same corral setup as in Fig.
\ref{fig3}, the calculated LDOS at the center $\boldsymbol{r}\mathtt{=}0$ are
shown in Fig. \ref{fig4}. From Fig. \ref{fig4} one can find three prominent
features as follows: (i) the NM corral will induce an obvious bound state peak
(red curve) at the Dirac point; (ii) the AFM corral has little influence
(green dotted curve); (iii) the FM corral enhances the interaction between the
two magnetic impurities (blue curve) along with a number of resonance peaks
and an opened energy gap across the Dirac point. Therefore, we can conclude
that both NM and FM corrals (rather than AFM ones) on the TI surface may be
used to manipulate the interaction between magnetic impurities. This is
different from that in superconductors \cite{Morr} in which the quasiparticle
resonance is sensitive to the AFM corral.

Recently, some STM experiments about the impurity and defects on Bi$_{1-x}%
$Sb$_{x}$ \cite{Roushan}, Bi$_{2}$Te$_{3}$ \cite{Alpichshev, XueQK}, and
Bi$_{2}$Se$_{3}$ \cite{Alpichshev2} have been performed and resonance states
at Dirac point have been observed in Bi$_{2}$Se$_{3}$ \cite{Alpichshev2},
where the scattering potential strength is consistent with the parameters used
herein. To experimentally verify our predictions, the spin-polarized STM tip
and strong scattering potentials are required, which we believe are achievable
in current experimental capabilities, thereby, we hope our findings could be
observed in future experiments.

Compared to an ordinary (e.g. narrow gap) insulator or metal, the consequence
of the TI that we show in this paper is its unique quasiparticle interference
states confined in a corral. For example, as shown in Figs. 2(c) and 2(d), the
quasiparticle resonance states in a FM corral is highly spin resolved in
energy, which fundamentally distinguish the present corral-TI system from
previously reported ones harnessed on the ordinary metals or semiconductor
heterostructures. These novel phenomena are closely associated with the
topological chiral spin structure for TI surface states of Dirac type, and are
totally absent in an ordinary insulator or metal, in which the quasiparticle
interference states in a quantum corral have been extensively studied.
Different from the normal metal surface states, the Dirac point in TI surface
states is so stable against impurity scattering that the weak potential
impurities cannot induce significant modification in LDOS near the Dirac
point, such as resonance states at Dirac point. Whereas, the low-energy
resonance states due to the impurity scattering on ordinary insulator or metal
surface are easily to be observed with weaker potential impurity \cite{Lobos}.
Moreover, even if the Rashba SOI in metal surface states is taken into
account, the main features of quantum corrals on normal metal Au(111) surface
is similar to those without SOI since the Rashba SOI is much weaker than the
kinetic energy term in ordinary insulator or metal. 

Besides, due to the single Dirac point nature, no complicated intervalley
scattering events occur on TI surface. Therefore, it can be expected that
\textit{in situ} measurement or even manipulation of impurity scattering
processes on TI surfaces should pave a promising way to study electronic
structures and find extraordinary quantum phenomena in various TI materials.
The effects observed above, especially induced by FM corral, may provide a
useful method to control the electron's spin, and provides a useful criterion
to distinguish the aligned mode of the magnetic corral on TI surface from
other 2D systems with an even number of Dirac points, such as graphene. In
graphene, the two components of the low-energy Hamiltonian indicate the
pseudospin of the two sublattices not real spin that link to the quasiparticle
momentum, while the twofold spin degeneracy is preserved, and the spin does
not couple to magnetic impurities directly in graphene. Consequently, no net
spin-polarized LDOS or chirality will be observed in graphene even if the
intervalley scattering is ignored.

\section{Conclusions}

In summary, we have studied electron interference in the LDOS induced by NM,
AFM, and FM quantum corrals on a TI surface. We have found that the impurity
resonance states are much easier to occur in a FM corral, in which the
resonant peaks become much sharper with increasing the magnetic impurity
potentials, so that the LDOS is suppressed and an energy gap at the Dirac
point is opened. Especially, the spin-up and spin-down LDOSs are well
energy-resolved in a FM corral, which may be useful for spin selection and
developing spintronics devices in TI. Furthermore, a quantum mirage of a
magnetic impurity bound state is projected in a FM elliptic corral, while for
the NM or AFM corrals it is too weak to be observed. In addition, we have also
discussed the interaction between two magnetic impurities in quantum corrals.
Interestingly, this interaction is prominently enhanced in NM and FM corrals,
rather than in an AFM one. These effects may offer new guide in manipulating
topological surface states, as well as potential applications for spintronics
and quantum information.

The authors would like to thank Jamie D. Walls for helpful suggestions on the
manuscript. This work was supported by NSFC under Grants No. 90921003, No.
60776063, No. 60821061, and 60776061, and by the National Basic Research
Program of China (973 Program) under Grants No. 2009CB929103 and No. G2009CB929300.

\end{document}